\documentclass[twocolumn,aps,showpacs,superscriptaddress]{revtex4} 
\usepackage{epsfig}
\usepackage{amssymb}

\begin{document} 

\newcommand{\vk}{{\vec k}} 
\newcommand{\vK}{{\vec K}}  
\newcommand{\vb}{{\vec b}}  
\newcommand{\vp}{{\vec p}}  
\newcommand{\vq}{{\vec q}}  
\newcommand{\vQ}{{\vec Q}} 
\newcommand{\vx}{{\vec x}} 
\newcommand{\vh}{{\hat{v}}} 
\newcommand{\cO}{{\cal O}}
\newcommand{\tr}{{{\rm Tr}}}  
\newcommand{\be}{\begin{equation}} 
\newcommand{\ee}{\end{equation}}  
\newcommand{\half}{{\textstyle\frac{1}{2}}}  
\newcommand{\gton}{\mathrel{\lower.9ex \hbox{$\stackrel{\displaystyle 
>}{\sim}$}}}  
\newcommand{\lton}{\mathrel{\lower.9ex \hbox{$\stackrel{\displaystyle 
<}{\sim}$}}}  
\newcommand{\ben}{\begin{enumerate}}  
\newcommand{\een}{\end{enumerate}} 
\newcommand{\bit}{\begin{itemize}}  
\newcommand{\eit}{\end{itemize}} 
\newcommand{\bc}{\begin{center}}  
\newcommand{\ec}{\end{center}} 
\newcommand{\bea}{\begin{eqnarray}}  
\newcommand{\eea}{\end{eqnarray}}

\title{Elliptic flow at large transverse momenta from quark coalescence}
 
\date{\today}
 
\author{D\'enes Moln\'ar}
\affiliation{Department of Physics, Ohio State University,
                174 West 18th Ave, Columbus, OH 43210}
\author{Sergei A. Voloshin}
\affiliation{Department of Physics and Astronomy, Wayne State
University, 666 W. Hancock, Detroit, MI 48201}

\begin{abstract}
We show that hadronization via quark coalescence 
enhances hadron elliptic flow at 
large $p_\perp$ relative to that of partons
at the same transverse momentum. 
Therefore, compared to 
earlier results based on covariant parton transport theory,
more moderate initial parton densities $dN/d\eta(b=0) \sim 1500-3000$ 
can explain the differential elliptic flow $v_2(p_\perp)$ data 
for $Au+Au$ reactions at $\sqrt{s}=130$ and $200 A$ GeV from RHIC.
In addition, $v_2(p_\perp)$ could saturate 
at about 50\% higher values for baryons than for mesons.
If strange quarks have weaker flow than light quarks,
hadron $v_2$ at high $p_\perp$ decreases with relative strangeness
content.

\end{abstract}


\maketitle 

{\em Introduction.}
The goal of relativistic  heavy ion collision experiments is to produce
{\em macroscopic} amounts of deconfined partonic matter and study its collective behavior.
One of the important experimental probes of collective
dynamics in $A+A$ reactions is 
differential elliptic flow\cite{flow-review}, $v_2(p_\perp)\equiv\langle \cos(2\phi)\rangle_{p_\perp}$,
 the second Fourier moment of the azimuthal momentum distribution
for a given $p_\perp$.
Measurements of elliptic flow at high transverse momentum
provide important constraints
about the density and effective energy loss of partons\cite{pQCDv2,v2}.

Recent data from RHIC for Au+Au reactions at $\sqrt{s_{NN}}=130$ and $200$~GeV
show a remarkable saturation property of elliptic flow 
in the region
$2$ GeV $< p_\perp < 6$ GeV with $v_2$ 
reaching up to 0.2 \cite{PHENIXv2,STARv2,KirillQM2002,PHENIXidentv2}.
The saturation pattern, 
which corresponds to a factor of two azimuthal angle asymmetry 
of  high-$p_\perp$ particle production relative to the reaction plane,
is still waiting for theoretical explanation.

The saturation and eventual decrease 
of $v_2$ at high $p_\perp$ has been demonstrated 
as a consequence of finite inelastic parton energy loss\cite{pQCDv2}. 
Though the qualitative features in the data were explained,
for realistic diffuse nuclei
the calculations show a rapid {\em decrease} of 
$v_2$ above $p_\perp > 3-4$ GeV 
contrary to the saturation out to $p_\perp \approx 6$ GeV seen in the data.

Calculations of elliptic flow based on
ideal (nondissipative)
hydrodynamics\cite{Ollitrault:1992bk,Heinzhydro,Kolbhydro,Teaneyhydro}
can reproduce the low $p_\perp<2$ GeV data at RHIC remarkably well,
however overshoot the data above $p_\perp>2$ GeV \cite{STARv2,Kolbhighptv2}.
The lack of saturation is due to
the assumption of zero mean free path and
that local equilibrium can be maintained
throughout the evolution\cite{v2,Teaneyv2}.

Covariant parton transport theory\cite{ZPC,ZPCv2,nonequil,inelv2,v2,hbt}
overcomes this problem via replacing
the assumption of local equilibrium by
that of a finite local mean free path $\lambda(s,x) \equiv 1/\sigma(s) n(x)$. 
The theory then naturally  
interpolates between  free streaming ($\lambda=\infty$)
and  ideal hydrodynamics ($\lambda=0$).
Several studies confirm
that initial parton densities\cite{EKRT}
and elastic $2\to 2$ parton cross sections estimated from perturbative QCD,
$dN_g/d\eta(b=0) \sim 1000$ and $\sigma_{gg\to gg}\approx 3$ mb,
generate too small collective effects at RHIC\cite{ZPCv2,inelv2,v2,hbt}.
Nevertheless, quantitative agreement with the $v_2(p_\perp)$ data 
is possible, provided initial parton densities and/or cross sections
are enhanced by an order of magnitude
to $\sigma dN_g/d\eta(b=0) \sim 45000$ mb\cite{v2}.
A similar enhancement is indicated by the pion interferometry data
as well\cite{hbt}.
The origin of such an opaque parton environment is the RHIC ``opacity puzzle''.

To compare to the experiments,
parton transport models also have to incorporate 
the hadronization process.
The studies mentioned above considered two simple schemes:
$1 parton \to 1\pi$ hadronization, motivated by parton-hadron duality,
and independent fragmentation.
An alternative model of hadronization is quark coalescence,
in which the relevant degrees of freedom 
are not free partons but massive (dressed) valence quarks. 
Gluons are assumed to have 
converted to quarks, therefore there are no dynamical gluons considered.

Quark coalescence
has been applied successfully
in the ALCOR\cite{ALCOR} and MICOR\cite{MICOR} 
models to explain particle abundances and spectra in heavy-ion collisions.
It was also suggested recently in Ref. \cite{Voloshincoal} 
as an explanation for the anomalous meson/baryon ratio and 
features of the elliptic flow data  at RHIC.
In this letter we 
show that hadronization via quark coalescence
can resolve most of the ``opacity puzzle''
because it leads to an amplification of elliptic flow at high $p_\perp$.

{\em Quark coalescence.}
The usual starting point of coalescence models is
the statement that the invariant spectrum of produced particles is
proportional to the {\em product} of the invariant spectra of 
constituents\cite{coal1}.
This means that 
(assuming that different quark and anti-quark distributions are the
same) the hadron spectra at midrapidity  
are given by those of partons via
\bea
\frac{dN_B}{d^2 p_\perp}(\vp_\perp) 
 &=& C_B(p_\perp) \left[  \frac{dN_q}{d^2 p_\perp}(\vp_\perp/3) \right]^3
\nonumber\\
\frac{dN_M}{d^2 p_\perp}(\vp_\perp) 
 &=& C_M(p_\perp) \left[  \frac{dN_q}{d^2 p_\perp}(\vp_\perp/2) \right]^2 \ ,
\label{coal_eq}
\eea
where the coefficients 
$C_M$ and $C_B$ 
are the probabilities for
$q\bar{q}\to meson$ and $qqq \to baryon$ coalescence.
We allow for $p_T$ dependent coalescence factors
because more careful treatment of the coalescence problem\cite{coal2}
shows that such a dependence may arise, e.g., 
due to kinematic (energy) factors or strong radial flow. 
This, however, does not influence elliptic flow because it is a ratio from 
which the coalescence factors drop out (see Eq.~(\ref{v2})).

These relations are only valid for rare processes. 
This is {\em not} the case at high constituent phase space densities,
when most quarks recombine into hadrons and 
hence the number of hadrons is {\em linearly} proportional to 
that of quarks, $dN_h(p_\perp) \propto dN_q(p_\perp)$.

At lower constituent densities, coalescence processes
become relatively rare and 
therefore the usual coalescence formalism works.
On the other hand, most quarks hadronize via fragmentation into hadrons.
Nevertheless, depending on how quickly the parton phase space density
drops with increasing $p_\perp$,
there can be a region of {\em hadron} transverse momenta 
that is populated dominantly via coalescence.
The reason for this is that
hadrons from coalescence have {\em larger} momenta than the average quark
momentum,
$dN_h^{coal}(p_\perp) = C_h [dN_q(p_\perp/n)]^n$, $(n=2,3)$,
whereas hadrons from fragmentation 
carry only a {\em fraction} $z < 1$ of the initial quark momentum,
$dN_h^{frag}(p_\perp) \sim dN_q(p_\perp/z)$.

At very low parton densities, e.g., at very high transverse momentum,
the fragmentation process wins,
in accordance with the QCD factorization theorem.
For example, a power law parton 
spectrum $dN_q/p_\perp dp_\perp \sim A p_\perp^{-\alpha}$
implies $dN_h^{coal}/dN_h^{frag} \sim C_h A^{n-1} p_\perp^{-(n-1)\alpha}
\to 0$ at high $p_\perp$.

Therefore,
in heavy-ion collisions there can be
three qualitatively different phase space regions.
At very large
transverse momenta particle productions is dominated by independent parton 
fragmentation.
At lower transverse momenta coalescence prevails,
which region can itself be subdivided into two parts:
a very low $p_\perp$ (high phase space density) region where
Eq. (\ref{coal_eq}) is {\em not} applicable, 
and a moderate density (higher $p_\perp)$
region, where Eq. (\ref{coal_eq}) is valid.
Because the density of produced particles depends on
the centrality of the collision,
the ``boundaries'' of these regions depend on centrality. 
Only detailed quantitative studies\cite{MullerMtoB,LevaiMtoB}
of the relative contributions of the various hadronization processes,
which is beyond the scope of this letter,
could determine where the exact bounds are.
Alternatively, the limits can be deduced from comparison
with the experimental data.

{\it Anisotropic flow.}
For brevity we discuss only elliptic flow as the most important and
interesting case.
However, Eqs.~(\ref{v2LO}), (\ref{v2LOdiff}),
and all conclusions below also apply
(i) when azimuthal anisotropies 
$v_{k}(p_\perp)\equiv \langle \cos(k\phi)\rangle_{p_\perp}$ of any order
are present,
and (ii) to any anisotropy coefficient $v_{k}$ instead of $v_2$, 
even in the former most general case.

In the coalescence region, meson and baryon elliptic flow are given
by that of partons via
\be
v_{2,M}(p_\perp) \approx  2 v_{2,q} (\frac{p_\perp}{2}) , \quad
v_{2,B}(p_\perp) \approx  3 v_{2,q} (\frac{p_\perp}{3}) \ ,
\label{v2LO}
\ee
as follows from Eq.~(\ref{coal_eq}) and $v_2 \ll 1$.
For example, if partons have only elliptical anisotropy,
i.e., $dN_q/p_\perp dp_\perp d\Phi = (1/2\pi) dN_q/p_\perp dp_\perp
[1+2v_{2,q} \cos (2\Phi)]$, then
\bea
 v_{2,B}(p_\perp)  &=& \frac{3 v_{2,q} (p_\perp/3) + 3 v_{2,q}^3(p_\perp/3)}{1+6 v_{2,q}^2 (p_\perp/3)}
 \nonumber \\
 v_{2,M}(p_\perp)  &=& \frac{2 v_{2,q} (p_\perp/2)}{1+2 v_{2,q}^2 (p_\perp/2)} 
\ .
\label{v2}
\eea

\begin{figure}[hbpt] 
\hspace*{-0.2cm}\epsfig{file=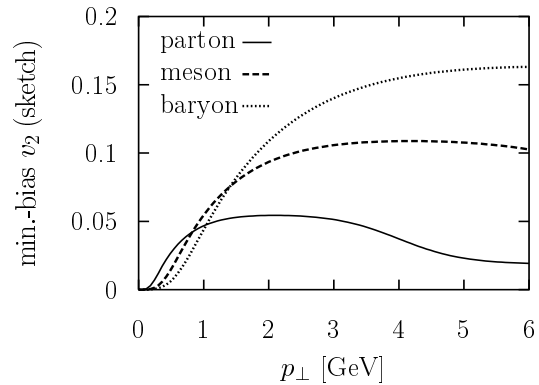,height=2.0in,width=2.8in,clip=5,angle=0} 
\begin{minipage}[t]{8.6cm}  
\vspace*{-0.5cm} 
\caption{\label{fig1} 
Qualitative behavior of baryon and meson elliptic flow as a function of $p_\perp$ from quark coalescence.
}
\end{minipage} 
\end{figure}

Fig.~\ref{fig1} illustrates the effect of quark coalescence on 
baryon and meson elliptic flow compared to parton elliptic flow.
The latter is shown schematically by the solid line.
At small transverse momenta, parton $v_2(p_\perp)\propto p_\perp^2$, 
as follows from general analyticity considerations. 
This region, before $v_2$ becomes approximately linear in $p_\perp$ 
could be relatively small
(depending on the effective mass of partons).
At higher transverse momenta $p_\perp > 1-2$~GeV,   
parton elliptic flow saturates as predicted by parton transport\cite{v2},  
and then, possibly already above $p_\perp \gton 4$ GeV, 
decreases according to pQCD parton energy loss calculations\cite{pQCDv2}.  
The curve for baryon(meson) elliptic flow has been obtained by
simply rescaling the parton curve by a factor three(two) both vertically and
horizontally. 
As discussed above, for very low and very high $p_\perp$, 
we boldly use Eq.~(\ref{v2LO}) beyond its
region of applicability but doing so 
does not affect the discussion.

There are three qualitatively different regimes in Fig.~\ref{fig1}:
 
\noindent
(i) In the small $p_\perp$ region where $v_2(p_\perp)$ increases
faster than linearly, $v_{2,B}<v_{2,M}< v_{2,q}$. 
It is not clear to what extent
the coalescence picture is applicable in this region
but it is interesting that the data does exhibit such a behavior.
This ordering follows naturally from hydrodynamics,
where flow decreases with increasing
particle mass\cite{Heinzhydro,Kolbhydro,Teaneyhydro}.
Similar mass dependence could also arise in a coalescence model
because heavier hadrons can be formed by quarks with
larger relative momentum (ignored in the current approach).

\noindent
(ii) In the intermediate $p_\perp$ 
region where $v_2(p_\perp)$ depends linearly on transverse
momentum, $v_{2,B} \approx v_{2,M}$.

\noindent
(iii) At large $p_\perp$, where parton
$v_2(p_\perp)$ increases slower than linearly, 
baryon flow becomes larger than that of mesons,
$v_{2,B}>v_{2,M} > v_{2,q}$, by as much as 50\%.    
Parton collective flow saturation,
predicted for $p_\perp > 1-2$~GeV by parton transport\cite{v2},
results in saturating meson/baryon flow at  $p_\perp > 2-4$~GeV  that is
{\em amplified two/three-fold}  compared to that of partons.
Saturation sets in at $50\%$ higher $p_\perp$ for 
baryons than for mesons.
In addition, any eventual decrease of parton elliptic flow
at very high $p_\perp$,
would happen at two to three times larger $p_\perp$ for hadrons.

The high-$p_\perp$ results above strongly differ from those 
obtained in Ref.~\cite{LinKov2}.
The reason is that, unlike Eq.~(\ref{v2LO}), in Ref.~\cite{LinKov2} the
coalescence of quarks was considered to be independent of their relative
momenta and therefore hadron elliptic flow at high
$p_\perp$ was similar to that of a high-$p_\perp$ quark.   

If not all quarks show the same elliptic flow, further
 differentiation occurs because in that case
\bea
v_{2,B=abc}(p_\perp) &\approx& v_{2,a}(p_\perp/3) + v_{2,b}(p_\perp/3) +v_{2,c}(p_\perp/3)
\nonumber\\
v_{2,M=\bar{a}b}(p_\perp) &\approx& v_{2,\bar{a}}(p_\perp/2) + v_{2,b}(p_\perp/2)\ .
\label{v2LOdiff}
\eea
For example, strange quarks may have a smaller $v_2(p_\perp)$ 
than light quarks, at high $p_\perp$ because heavy quarks are expected
 to lose less energy 
in nuclear medium\cite{heavyqEL},
while at low $p_\perp$ due to the mass dependence of hydrodynamic flow.
If $v_2^s < v_2^q$, elliptic flow decreases with increasing relative 
strangeness content within the baryon and meson bands,
i.e., $v_2^p > v_2^\Lambda \approx v_2^\Sigma > v_2^\Xi > v_2^\Omega$ 
and $v_2^\pi > v_2^K > v_2^\phi$.

{\em Possible solution to the opacity puzzle.}
While hadronization via $1parton \to 1\pi$ or independent fragmentation
approximately preserves elliptic flow at high $2 < p_\perp < 6$ GeV 
\cite{v2},
quark coalescence enhances $v_2$ two times for mesons and three times for 
baryons.
Hence, the same hadron elliptic flow can be reached
from $2-3$ times smaller parton $v_2$,
which requires smaller parton opacities, i.e., initial parton 
densities and/or cross sections.
The amplification also allows the RHIC $v_2$ data to exceed
geometric upper bounds derived based on a nuclear absorption 
model\cite{shuryaklimit}
(the data are compatible with those constraints only 
for idealistic ``sharp sphere'' nuclear distributions\cite{Voloshincoal}).
Those bounds apply to the parton $v_2$
and thus are two/three times higher for mesons/baryons.

To determine the reduction of parton opacity quantitatively,
we rely on the results of Ref.~\cite{v2} that
computed gluon elliptic flow as a function of 
the transport opacity, 
$\chi\equiv \int dz\, \sigma_{tr} \rho(z) 
\approx \sigma dN/d\eta\,/\, (940$ mb$)$, from 
elastic parton transport theory for a minijet scenario of Au+Au at 
$\sqrt{s}= 130A$ GeV at RHIC.
Those results can be conveniently parameterized as 
$v_2(p_\perp,\chi) = v_2^{max}(\chi) \tanh[p_\perp / p_0(\chi)]$,
where $v_2^{max}$ is the saturation value of elliptic flow,
while $p_0$ is the $p_\perp$ scale above which saturation sets in.
For the estimates here 
we assume that all gluons convert, e.g., via $gg\to qq$, 
to quarks of similar $p_\perp$ and hence $v_2^q(p_\perp) = v_2^g(p_\perp)$.

\begin{figure}[hbpt] 
\hspace*{-0.2cm}\epsfig{file=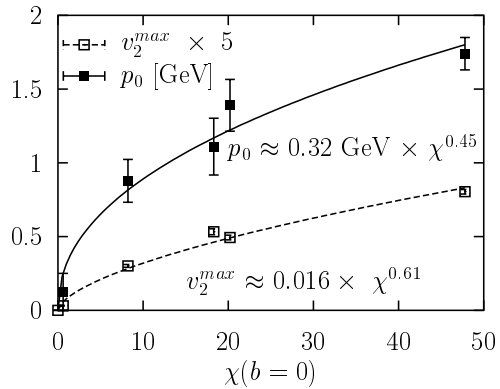,height=2.05in,width=2.6in,
clip=5,angle=0} 
\begin{minipage}[t]{8.6cm}  
\vspace*{-0.5cm} 
\caption{\label{fig2} 
Fit parameters $v_2^{max}$ and $p_0$ as a function of transport
opacity $\chi$(b=0), for the $v_2(p_\perp)$ results in Ref.~\cite{v2}.
}
\end{minipage} 
\end{figure} 
As shown in Fig.~\ref{fig2}, 
the increase of elliptic flow with opacity is weaker than linear, 
$v_2^{max}\sim \chi^{0.61}$.
Therefore, a $2-3$ times smaller
parton elliptic flow,
which is needed to match the charged particle $v_2$ data from RHIC
in our coalescence scenario,
corresponds to $3-6$ times smaller parton opacities 
$\sigma dN/d\eta(b=0) \sim 7000-15000$ mb than those found in Ref.~\cite{v2}.
The lower(upper) value applies if high-$p_\perp$ hadrons are mostly
 baryons(mesons). 
Based on preliminary PHENIX data \cite{PHENIXhighpt} showing 
$\pi_0 / h^{\pm} \approx 0.5$ between $2 < p_\perp < 9 $ GeV,
one may expect $mesons/baryons\approx 1$,
in which case $\sigma dN/d\eta(b=0) \approx 10000$~mb.

In Ref.~\cite{v2} only collective flow was considered and
the parton opacity at RHIC was extracted using elliptic flow data from the
reaction plane analysis.
Taking into account non-flow effects that contributed
up to 15-20\% \cite{cumulantv2}
to the first elliptic flow measurements,
parton opacities should be further reduced by $25\%$
to $\sigma dN/d\eta(b=0) \sim 5000-10000$~mb.
For a typical elastic $gg\to gg$ cross section of $3$ mb,
this corresponds to an initial parton density $dN/d\eta(b=0) \sim 1500-3000$,
only $1.5-3$ times above the EKRT perturbative estimate\cite{EKRT}.

The remaining much smaller discrepancy is comparable to
theoretical uncertainties.
For example, perturbative cross section and parton density
estimates may be too low.
If most hadrons formed via coalescence,
the observed hadron multiplicity $dN_h/d\eta \approx 1000$ would imply
much higher initial parton densities $dN/d\eta \sim 2000-3000$.
Constituent quark cross sections, $\sigma_{qq} \approx 4-5$ mb,
also point above the $\approx 3$ mb perturbative estimate.
One effect that estimate ignores is the enhancement of parton cross sections 
$\sigma \propto \alpha_s^2/\mu^2$
due to the decrease 
of the self-consistent Debye screening mass 
$\mu \sim gT_{eff}(\tau)$ during the expansion.
Finally, the contribution of inelastic processes,
such as $gg\leftrightarrow ggg$, to the opacity has also 
been neglected so far.
A preliminary study shows\cite{inelv2} that
this contribution can be similar to that of elastic processes.

{\em Summary.}
In this letter we studied elliptic flow of hadrons formed 
from coalescence of quarks with similar momenta.
At high $p_\perp > 2$ GeV we found an enhancement of elliptic flow 
compared to that of partons.
With the enhancement,
moderate initial parton densities $dN_g/d\eta \sim 1500-3000$
are sufficient to account for the charged particle elliptic flow 
data from RHIC, providing a possible solution to the RHIC ``opacity puzzle''.
At low $p_\perp < 1$ GeV, on the other hand,
hadron elliptic flow is suppressed.

Quark coalescence gives a weaker baryon flow than meson flow
at low $p_\perp < 0.5-1$ GeV,
while the opposite, $v_2^B > v_2^M$, at high $p_\perp > 2-3$ GeV.
Assuming all partons have similar elliptic flow,
$v_2^B \approx 1.5 v_2^M$ at high $p_\perp$.
If on the other hand strange quarks show weaker flow than light quarks,
elliptic flow at high $p_\perp$ is ordered 
by relative strangeness content, such that
$v_2^p > v_2^\Lambda \approx v_2^\Sigma > v_2^\pi > v_2^K > v_2^\phi$,
$v_2^{\Lambda,\Sigma} > v_2^\Xi > v_2^K$, and $v_2^\Xi > v_2^\Omega 
\approx 3 v_2^\phi/2$. 
These predictions can be readily tested in current and future heavy-ion 
collision experiments.

We emphasize that the quark coalescence picture and therefore our 
flow ordering predictions 
strongly rely on the assumption that quark degrees
of freedom are dominant at hadronization. Therefore, experimental support for 
our predictions may indicate the formation of deconfined nuclear matter 
in heavy ion collisions at RHIC energies.

We also note that at very high $p_\perp$ one expects 
a transition from hadronization 
via quark coalescence to independent fragmentation. 
An experimental signature of this may be the 
reduction of baryon $v_2$ below meson $v_2$.

When this work was in its final stage,
two preprints
addressing baryon to meson ratio at high $p_\perp$,
\cite{MullerMtoB} and \cite{LevaiMtoB}, were submitted
to the {\sf arXiv.org} e-print server.
While these studies mainly focus on baryon and meson yields,
the underlying physical arguments are very similar to those presented here.

{\em Acknowledgments:} 
Valuable comments by U.~Heinz and M.~Gyulassy
are gratefully acknowledged. 
This work was supported 
by DOE grants DE-FG02-01ER41190 and DE-FG02-92ER40713.


\begin{thebibliography}{99}
 


\bibitem{flow-review}
For a recent review see, 
e.g.:
J.~Ollitrault, Nucl. Phys. A {\bf 638}, 195 (1998); 
A.~M.~Poskanzer,
nucl-ex/0110013;
or Ref.~\cite{Voloshincoal}.

\bibitem{pQCDv2}
X.~Wang,
Phys.\ Rev.\ C {\bf 63}, 054902 (2001);
M.~Gyulassy, I.~Vitev and X.~N.~Wang,
Phys. Rev. Lett. {\bf 86}, 2537 (2001);
M.~Gyulassy {\it et al.},
Phys.\ Lett.\ B {\bf 526}, 301 (2002).

\bibitem{v2}
D.~Molnar and M.~Gyulassy,
Nucl.\ Phys.\ A {\bf 697}, 495 (2002);
Erratum-{\it ibid} A {\bf 703}, 893 (2002).

\bibitem{STARv2}
C.~Adler {\it et al.}  [STAR Collaboration],
Phys. Rev. Lett. 90, 032301 (2003).
{\it ibid} {\bf 89}, 132301 (2002);
{\it ibid}  {\bf 87}, 182301 (2001).

\bibitem{PHENIXv2}
K.~Adcox  [PHENIX Collaboration],
Phys.\ Rev.\ Lett.\  {\bf 89}, 212301 (2002).

\bibitem{PHENIXidentv2}
S.~Esumi  [PHENIX Collaboration],
Nucl.\ Phys.\ A{\bf 715}, 599 (2003).

\bibitem{KirillQM2002}
K.~Filimonov  [STAR Collaboration],
Nucl.\ Phys.\ A{\bf 715}, 737 (2003).

\bibitem{Ollitrault:1992bk}
J.~Ollitrault,
Phys.\ Rev.\ {\bf D 46}, 229 (1992).

\bibitem{Heinzhydro}
P.~F.~Kolb, J.~Sollfrank and U.~W.~Heinz,
Phys.\ Lett.\ B {\bf 459}, 667 (1999);
Phys.\ Rev.\ C {\bf 62}, 054909 (2000).

\bibitem{Teaneyhydro}
D.~Teaney, J.~Lauret and E.~V.~Shuryak,
Phys.\ Rev.\ Lett.\  {\bf 86}, 4783 (2001).

\bibitem{Kolbhydro}
P.~F.~Kolb {\it et al.},
Nucl.\ Phys.\ A {\bf 696}, 197 (2001);
Phys.\ Lett.\ B {\bf 500}, 232 (2001);
P.~Huovinen {\it et al.},
Phys.\ Lett.\ B {\bf 503}, 58 (2001).

\bibitem{Kolbhighptv2}
U.~W.~Heinz and P.~F.~Kolb,
hep-ph/0204061.

\bibitem{Teaneyv2}
D.~Teaney,
Nucl.\ Phys.\ A{\bf 715}, 817 (2003).


\bibitem{ZPC}
B.~Zhang,
Comput.\ Phys.\ Commun.\  {\bf 109}, 193 (1998).

\bibitem{ZPCv2}
B.~Zhang, M.~Gyulassy and C.~M.~Ko,
Phys.\ Lett.\ {\bf B455}, 45 (1999).

\bibitem{inelv2}
%
D.~Moln\'ar, Nucl.\ Phys. {\bf A661}, 236 (1999).

\bibitem{nonequil}
D.~Molnar and M.~Gyulassy,
Phys.\ Rev.\ C {\bf 62}, 054907 (2000).

\bibitem{hbt}
D.~Molnar and M.~Gyulassy,
nucl-th/0211017.

\bibitem{EKRT}
K.~J.~Eskola {\it et al.},
Nucl. Phys. {\bf B570}, 379 (2000).

\bibitem{ALCOR}
T.~S.~Biro, P.~Levai and J.~Zimanyi,
Phys.\ Lett.\ B {\bf 347}, 6 (1995).

\bibitem{MICOR}
P.~Csizmadia and P.~Levai,
J.\ Phys.\ G {\bf 28}, 1997 (2002).

\bibitem{coal1}
A.~Schwarzschild and C.~Zupancic,
Phys.\ Rev.\ {\bf 129}, 854 (1963);
H.~Sato and K.~Yazaki,
Phys.\ Lett.\ B {\bf 98}, 153 (1981).

\bibitem{coal2}
S.~T.~Butler and C.~A.~Pearson,
Phys.\ Rev.\ {\bf 129}, 836 (1963);
C.~B.~Dover  {\it et al.}, 
Phys.\ Rev.\ C {\bf 44}, 1636 (1991);
R.~Scheibl and U.~W.~Heinz,
Phys.\ Rev.\ C {\bf 59}, 1585 (1999).


\bibitem{Voloshincoal}
S.~A.~Voloshin,
Nucl.\ Phys.\ A{\bf 715}, 379 (2003).

\bibitem{LinKov2}
Z.~w.~Lin and C.~M.~Ko,
Phys.\ Rev.\ Lett.\  {\bf 89}, 202302 (2002).

\bibitem{heavyqEL}
M.~H.~Thoma and M.~Gyulassy,
Nucl.\ Phys.\ B {\bf 351}, 491 (1991);
E.~Braaten and M.~H.~Thoma,
Phys.\ Rev.\ D {\bf 44}, 2625 (1991);
Y.~L.~Dokshitzer and D.~E.~Kharzeev,
Phys.\ Lett.\ B {\bf 519}, 199 (2001).

\bibitem{shuryaklimit}
E.~V.~Shuryak,
Phys.\ Rev.\ C {\bf 66}, 027902 (2002).

\bibitem{PHENIXhighpt}
S.~Mioduszewski  [PHENIX Collaboration],
Nucl.\ Phys.\ A{\bf 715}, 199 (2003).

\bibitem{cumulantv2}
C.~Adler {\it et al.}  [STAR Collaboration],
Phys.\ Rev.\ C {\bf 66}, 034904 (2002).

\bibitem{MullerMtoB}
R.~J.~Fries {\it et al.},
Phys.\ Rev.\ Lett.\  {\bf 90}, 202303 (2003).

\bibitem{LevaiMtoB}
V.~Greco, C.~M.~Ko and P.~Levai,
Phys.\ Rev.\ Lett.\  {\bf 90}, 202302 (2003).

\end{thebibliography}
\end{document}